\documentstyle[aps,multicol,epsf]{revtex}
\input epsf
\draft
\title{Filtered Talbot lens: Producing $\lambda /2n$-periodic atomic patterns with
standing wave fields having period $\lambda $}
\author{J. L. Cohen, B. Dubetsky, and P. R. Berman 
\\ {\em Physics Department, University of Michigan, Ann Arbor, MI 48109-1120} }
\author{J. Schmiedmayer 
\\ {\em Institut f\"{u}r Experimentalphysik, Universit\"{a}t-Innsbruck,\\
Technikerstrasse 25, A-6020 Innsbruck, Austria} }
\date{\today  }
\begin{document}
\maketitle
\begin{abstract}
We propose a scheme to create high-contrast, periodic atom density
distributions having period $\lambda /2n$ using the Talbot effect, where $%
\lambda $ is the wave length of the optical fields that scatter the atoms
and $n$ is a positive integer. This {\em filtered Talbot lens} is comprised
of two standing-wave optical fields. An atomic beam propagates perpendicular
to the fields. The first field, which is far-detuned from the atomic
transition frequency, acts as an array of lenses that focuses the atoms. The
second field, positioned at the atom optical focus of the first, is resonant
with the atomic transition frequency and acts as an amplitude mask, leaving
unperturbed only those atoms that pass through its nodes. At distances
following the interaction with the second field that are equal to an
integral fraction of the Talbot length, atomic density gratings having
period $\lambda /2n$ are formed.
\end{abstract}


\begin{multicols}{2}
\narrowtext

\section{Introduction}

Over the past several years, considerable progress has been made in the
manipulation of atoms by optical fields. One important application of this
new technology is the focusing and deposition of atoms on substrates \cite
{focus,1}. Using standing wave optical fields having wavelength $\lambda $,
one has been able to write a periodic array of lines or dots having period $%
\lambda /2$. The atomic ''lines'' or ''dots'' themselves have widths $w$
that are very small compared with $\lambda /2$; widths (half width at half
maximum) as small as $6.5$ nm have been achieved \cite{2}. With such a
resolution, one can envision writing structures having periods as small as $%
\lambda /2n,$ where $n,$ the atomic grating order, can be as large as 
\begin{equation}
n_{\max }\sim \lambda /4w\sim 25.  \label{1}
\end{equation}
Although periods as small as $\lambda /2n_{\max }$ are consistent with the
resolution achieved to date, it is not obvious how one can use optical
fields having wavelength $\lambda $ to produce such periodic structures. For
example, focusing by standing wave fields results in periods equal to $%
\lambda /2$, or possibly $\lambda /4$ if cross-polarized counterpropagating
waves are used \cite{6}. A similar periodicity can be expected using optical
masks \cite{masks}, in which only those atoms passing through the nodes of a
standing wave field are registered on a substrate.

There have been a number of proposals for producing {\em sub-}$\lambda $ or 
{\em higher-order }atomic density gratings, which necessarily involve
nonlinear interactions of the atoms with the fields. Atom interference \cite
{3} allows one to exploit photon echo techniques to isolate higher order
gratings at specific focal planes following the interaction of an atomic
beam with two or more standing wave fields \cite{4,5}. High-order Bragg
scattering of an atomic beam by standing wave fields has produced structures
with periods as small as $\lambda /6$ \cite{7}. Counterpropagating waves,
detuned from one another by an appropriate ratio, can be used to create
arbitrarily high order gratings with good contrast \cite{highorder}. It is
even possible to achieve higher order gratings using the standard focusing
or optical mask geometries, if one is able to exploit the Talbot or
self-imaging effect \cite{talbot}

In atom optics the Talbot effect was observed using microfabricated
structures to scatter atoms \cite{8}. When atoms propagating along the $z$
axis pass through a microfabricated structure having period $\lambda /2$ in
the $x-$direction, the atomic spatial distribution becomes a periodic
function of $z$ having period equal to the Talbot length $L_{T}=\lambda
^{2}/2\lambda _{dB},$ where $\lambda _{dB}=h/Mu$ is the atomic de Broglie
wavelength, and $M$ and $u$ are atomic mass and speed, respectively. In
other words, self images of the microfabricated grating are produced at
integral multiples of the Talbot length. Moreover, for a microfabricated
structure with duty cycle $f$ (ratio of opening to period), periodic
structures having order as high as $\lambda /2f$ can be produced at integral
fractions of the Talbot length \cite{9}. Using this technique, 7th order
gratings have been observed \cite{nowak} using a metastable He beam passing
through a microfabricated structure having period $\lambda /2=6.55$ $\mu $
and duty cycle $f=0.1$.

\begin{figure}[tb!]
\centering
\begin{minipage}{8.0cm}
\epsfxsize= 8 cm \epsfysize= 8 cm \epsfbox{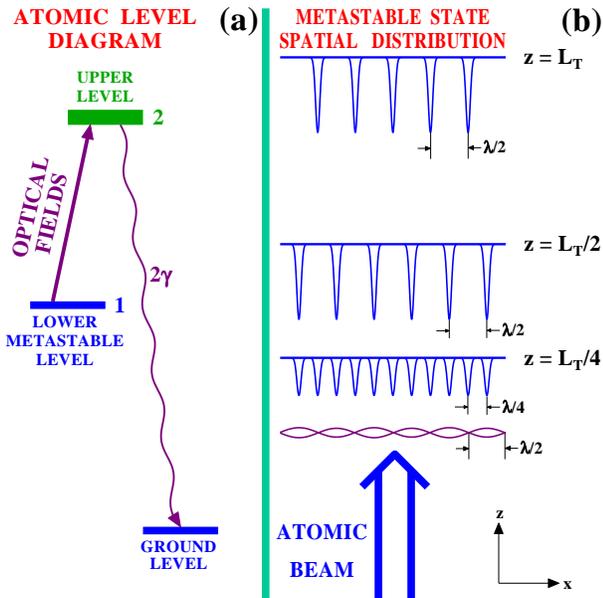}
\end{minipage}
\caption{Talbot effect and fractional self-images using an optical mask.
When a beam of metastable atoms having the level scheme shown in (a) passes
through the resonant standing wave field shown in (b), only atoms moving
within the narrow vicinity of the field nodes remain in level 1, producing a
metastable atomic density grating (not shown in the figure) having period $%
\protect\lambda /2$ immediately following the interaction. Further free
evolution of the atomic wave function results in a first-order self-image at
the Talbot distance $z=L_{T}$, a $\protect\lambda /4$ spatial shift of this
density at the distance $z=L_{T}/2$, a second-order self-image at the
distance $z=L_{T}/4$, and higher-order images at the fractional Talbot
distances $z=L_{T}/m$, for integer $m$.}
\label{fig1}
\end{figure}

Instead of microfabricated structures, one can use an optical mask to act as
an amplitude grating for a beam of metastable atoms. In a typical experiment 
\cite{masks}, a metastable atomic beam is sent through a standing wave field
that is resonant with a transition originating on the metastable level (see
Fig. \ref{fig1}). If the upper level is excited by the field, it decays to
the ground state. By using a substrate that is sensitive only to metastable
atoms, one exposes the substrate only to those atoms which pass through the
nodes of the field. Thus, for metastable atoms the strong standing wave
field acts as a microfabricated structure with a small duty cycle.
Experimentally, an atomic grating having period of order 200 nm ($\lambda
/4) $ was created using this effect when a metastable Ar beam was passed
through an optical mask \cite{schmied}. An alternative, non-lithographic
detection scheme that probes the lower level population may also be possible 
\cite{51}.

Optical masks give a high fringe contrast, but low deposition rate, since
most of the atoms pass through high field regions and are not recorded on
the substrate. To overcome this problem without an optical mask, one might
try to combine the Talbot effect with the focusing of atoms by a far-detuned
standing wave field. The field, which is also aligned along the $x$ axis,
acts as a series of lenses. The centers of the lenses are located at the
minima of the optical potential produced by the field, and the focal length
of each lens is denoted by $z_{f}$. If these focused atoms were equivalent
to atoms passing through an{\em \ amplitude transmission grating at the
focal plane,} one would expect that, as a result of the Talbot effect,
high-contrast, high-order images of the periodic focal pattern would be
formed at distances 
\begin{equation}
z_{m}=z_{f}+L_{T}/m.  \label{2}
\end{equation}
from the field. Janicke and Wilkens \cite{janicke} carried out numerical
calculations to obtain some evidence for this effect. A more detailed
analysis \cite{focusing}, however, reveals that {\em the background density
between the focused atoms seriously degrades the fractional Talbot effect}.

It might seem strange that the relatively small background density between
the focal spots can lead to a serious degradation of the fractional Talbot
signal. To understand the origin of this effect, it is helpful to return to
the theory of the fractional Talbot effect \cite{9}. If a wave function $%
\psi \left( x,z_{f}\right) $ is prepared at $z=z_{f}$ having period $d$,
then at a fractional Talbot distance $L_{T}/m$ from this plane, this wave
function is mapped into a set of $n$ copies shifted from one another along $x
$ by a distance $d/n$ $\left( \text{with }n=m\text{ or }m/2\right) .$ The
total wave function is a linear superposition of these copies with
coefficients having the same absolute value {\em but different phases}{\it . 
}After squaring the wave function to obtain the atomic density, one finds
that the different phases are responsible for the degradation of the
fractional Talbot images. In the case of focusing by an off-resonace field,
one finds that the tail of the wave function of one of the copies interferes
with the peak of the wave function of another copy. This is a strong effect.
A tail to peak ratio of 0.1 in the wave function at $z=z_{f}$ implies a
ratio of background to peak density of 0.01 at $z=z_{f}.$ However, at a
fractional Talbot distance from $z=z_{f}$, interference between the tail and
peak of the wave function can result in changes to the atomic density that
can deviate by as much as 20\% from the density that would have been
produced in the absence of any background density.

One way to avoid such interference effects is to minimize or eliminate the
background density. If a plane matter wave is incident on a microfabricated
grating or periodic amplitude mask having duty cycle $f,$ one period of the
wave function immediately following the mask consists of a peak having width 
$df$ and a vanishing background density having width $(1-f)d$. At a
fractional Talbot distance $L_{T}/m$ from this plane, there will be no
spatial overlap (and no interference) of the $n$ copies $\left( n=m\text{ or 
}m/2\right) $ of the wave function, provided that $n<1/f$. If this condition
is satisfied, a ''perfect'' fractional Talbot effect is produced.

To improve the quality of the fractional Talbot images, one can place a mask
in the path of the atoms just before they interact with the focusing field.
The mask reduces the aperture of each ''lens'' in the standing wave focusing
field, resulting in reduced spherical aberration and a reduction in
backgrond density \cite{icols}; however, just as in the case of optical
masks, a significant decrease in beam flux results. In this paper, we
combine off-resonant focusing with an optical mask placed at the focal plane
of the off-resonant field. The mask acts as a spatial filter for the focused
atoms that removes most of the background atoms between the focal spots.
Thus, the atomic beam flux is maintained while the degradation of the
fractional Talbot effect is avoided. As such, one can expect this {\em %
filtered Talbot lens }(FTL) to produce high density images of reduced
periodicity at integral fractions of the Talbot distance from the focal
plane. It is important to note that the wave fronts need not be planar at
the mask openings. As long as the waves incident on the {\em focusing} field
are nearly planar, the matter waves converging at the focal plane are
coherent. This coherent matter wave impinging on the mask produces a very
different diffracion pattern than that of an {\em incoherent}{\it \ }%
superposition of plane waves incident from different directions. An
incoherent superposition of waves would not produce a fractional Talbot
image on averaging over all directions of the incident wave, whereas a
coherent incident beam can produce a ''perfect'' fractional Talbot effect if
the duty cycle of the mask is sufficiently small.

We proceed below to examine this possibility in more detail.

\section{Filtered Talbot Lens}

A beam of atoms having the level structure shown in Fig. \ref{fig1}
interacts with two standing wave optical fields that drive the $1-2$
transition. The first field, located at $z=0,$ is far-detuned from the
atomic resonance and focuses the atoms at the plane $z=z_{f}$. The second
field, shown in Fig. \ref{fig1} and located at $z=z_{f}$, is resonant or
nearly resonant with the atoms. The atoms move with velocity ${\bf u}=u{\bf 
\hat{z}}$, and the $1-2$ transition frequency is denoted by $\omega $. In
the atomic rest-frame $\left( z=ut\right) $, the fields appear as two pulses
with electric field vectors given by 
\begin{eqnarray}
{\bf E}\left( x,t\right)  &=&{\bf \hat{y}}\frac{1}{2}\sum_{j=1,2}E_{j}e^{-i%
\Omega _{j}t}g_{j}\left( t-t_{j}\right)   \nonumber \\
&&\times \cos \left[ kx+\left( 1-j\right) \pi /2\right] +c.c.,  \label{3}
\end{eqnarray}
where $k=q/2$ is a propagation constant, $t_{1}=0,$ $t_{2}=t_{f}=z_{f}/u,$
and $E_{j}$ and $\Omega _{j}$ are the amplitude and frequency, respectively,
of the $j$th pulse. We assume that the pulse envelopes $g_{j}\left( t\right) 
$ are smooth functions of $t$ centered at $t=0,$ that both pulses have the
same duration $\tau $, and that $\tau \ll t_{f}$. {\em During} the
interaction with the $j$th pulse, the atomic state vector in the interaction
representation 
\[
{\bf \psi }\left( x,t\right) =\left( 
\begin{array}{c}
\psi _{2}\left( x,t\right)  \\ 
\psi _{1}\left( x,t\right) 
\end{array}
\right) 
\]
evolves as 
\begin{equation}
i\partial 
\bbox{\psi}
%
\left( x,t\right) /\partial t=\left( {\bf V}_{j}-i%
\bbox{\gamma}
%
\right) 
\bbox{\psi}
%
\left( x,t\right) ,  \label{4}
\end{equation}
and {\em between or after} the pulses the state vector evolves as 
\begin{equation}
i\partial 
\bbox{\psi}
%
\left( x,t\right) /\partial t=-\frac{\hbar }{2M}\frac{\partial ^{2}{\bf \psi 
}\left( x,t\right) }{\partial x^{2}}-i%
\bbox{\gamma}
%
\bbox{\psi}
%
\left( x,t\right) ,
\end{equation}
where $\hbar {\bf V}_{j}$ is the interaction Hamiltonian, 
\begin{equation}
{\bf V}_{j}{\bf =}\left( 
\begin{array}{cc}
0 & \chi _{j}g_{j}\left( t-t_{j}\right) e^{-i\Delta _{j}t} \\ 
\chi _{j}^{\ast }g_{j}^{\ast }\left( t-t_{j}\right) e^{i\Delta _{j}t} & 0
\end{array}
\right) \cos \left( kx\right) ,  \label{5}
\end{equation}
$\chi _{j}=-\mu E_{j}/2\hbar $ is a Rabi frequency, $\Delta _{j}=\Omega
_{j}-\omega $ is an atom-field detuning, and $\mu $ is a dipole moment
matrix element. The relaxation matrix 
\begin{equation}
\bbox{\gamma}
%
{\bf =}\left( 
\begin{array}{cc}
\gamma  & 0 \\ 
0 & 0
\end{array}
\right)   \label{6}
\end{equation}
corresponds to a case for which the lower state is metastable and the upper
state population decays with rate $2\gamma $ to level 0 {\em outside} of the 
$1-2$, two-level subspace. The decay rate of state $\left| 1\right\rangle $
has been set equal to zero under the assumption that its lifetime is much
greater than the Talbot time, 
\begin{equation}
T=L_{T}/u=\lambda ^{2}/\left( 2u\lambda _{dB}\right) =2\pi /\left( \hbar
q^{2}/2M\right) ,
\end{equation}
which is the relevant time scale in these focusing experiments. In Eq. (\ref
{4}) we have invoked the Raman-Nath approximation and neglected terms
associated with the kinetic energy of the atoms {\it during} the atom-field
interaction.

Before interacting with the first field, it is assumed that the collimated
beam can be approximated by a plane wave state in the $x$ direction having $%
x $-component of momentum $p=0$. If the atoms enter the first field zone in
state $\left| 1\right\rangle $, then the net effect of the far-detuned first
pulse $\left[ \left| \Delta _{1}\right| \gg \max \left( \left| \chi
_{1}\right| ,\gamma \right) \right] \ $is to produce a phase modulation of
the atomic lower state wave function, 
\begin{equation}
\psi _{1}\left( x,t=0^{+}\right) =\exp \left[ i\left( \theta /2\right) \cos
\left( qx\right) \right] ,  \label{8}
\end{equation}
where 
\begin{equation}
\theta =-\frac{\left| \chi _{1}\right| ^{2}}{\Delta _{1}}\int_{-\infty
}^{\infty }dt_{1}\left| g_{1}\left( t_{1}\right) \right| ^{2}  \label{9}
\end{equation}
is a pulse area. We assume that $\Delta _{1}<0$, resulting in $\theta >0.$

Following the action of the first pulse, the lower state wave function
evolves freely so that just before the second pulse acts, one has \cite
{focusing} 
\begin{equation}
\psi _{1}\left( x,t=t_{f}^{-}\right) =\sum_{s=-\infty }^{\infty
}i^{s}J_{s}\left( \theta /2\right) \exp \left( -is^{2}\omega
_{q}t_{f}+isqx\right) ,  \label{10}
\end{equation}
where $J_{s}\left( x\right) $ is a Bessel function of order $s\ $and 
\begin{equation}
\omega _{q}=\hbar q^{2}/2M=2\pi /T  \label{101}
\end{equation}
is a two-photon recoil frequency. When $\gamma t_{f}\gg 1$, \ one can
neglect the small upper state amplitude at $t=t_{f}.$ The atoms are assumed
to be focused at the time $t_{f}$ (distance $z_{f}=ut_{f})$ to lines
centered at $x=2\pi s/q$ for integer $s$.

To determine the effect of pulse 2, one must solve Eq. (\ref{4}) with $j=2$,
subject to the initial condition 
\begin{equation}
{\bf \psi }_{-}=\left( 
\begin{array}{c}
0 \\ 
\psi _{1}\left( x,t=t_{f}^{-}\right)
\end{array}
\right) \text{.}  \label{11}
\end{equation}
An analytic solution can be obtained if we assume that $\left| \dot{g}%
_{2}\left( t\right) /g_{2}\left( t\right) \right| \ll \gamma $ or 
\begin{equation}
\tau \gg \gamma ^{-1}.  \label{111}
\end{equation}
Condition (\ref{111}) is desirable as it guarantees that atoms will have
time to decay out of the two-state subspace during the interaction with the
second field pulse. Moreover, it is sufficient to satisfy the adiabatic
condition 
\begin{equation}
\tau \gg \min \left( \gamma ^{-1},\left| 2\chi _{2}\right| ^{-1},\left|
\Delta _{2}\right| ^{-1}\right) \text{.}  \label{adiabatic}
\end{equation}

As a result, the pulse turns on sufficiently slowly to ensure that
instantaneous eigenstates of the complex ''Hamiltonian'', $\hbar \left( {\bf %
V}_{2}-i%
\bbox{\gamma}
%
\right) $ are approximate eigenstates of the system. In this manner the
lower state wave function following the second pulse's action is given by 
\begin{equation}
\psi _{1}\left( x,t=t_{f}^{+}\right) \approx \eta (x)\psi _{1}\left(
x,t=t_{f}^{-}\right) ,  \label{12}
\end{equation}
where 
\begin{eqnarray}
\eta \left( x\right) =\exp \frac{i}{2}\int_{-\infty }^{\infty }dt_{1}\left[
\Delta _{2}+i\gamma \right.   \nonumber \\
-\left. \left( \left( \Delta _{2}+i\gamma \right) ^{2}+4\left| \chi
_{2}g_{2}\left( t_{1}\right) \right| ^{2}\sin ^{2}\left( kx\right) \right)
^{1/2}\right]   \label{121}
\end{eqnarray}
is the transmission function associated with the optical mask.

Following the second pulse, the lower state again evolves freely. When the
masking field is on exact resonance $\left( \Delta _{2}=0\right) $ and weak $%
\left( \left| \chi _{2}\right| /\gamma \ll 1\right) $, one finds 
\begin{equation}
\eta (x)=\exp \left[ -\beta \sin ^{2}\left( kx\right) \right] ,  \label{13}
\end{equation}
where 
\begin{equation}
\beta =\frac{\left| \chi _{2}\right| ^{2}}{\gamma }\int_{-\infty }^{\infty
}dt_{1}\left| g_{2}\left( t_{1}\right) \right| ^{2}  \label{14}
\end{equation}
is the area associated with pulse 2. Expanding expression (\ref{13}) in
plane waves and keeping in mind that each plane wave $e^{isqx}$ acquires the
time-dependent phase factor $e^{-is^{2}\omega _{q}\left( t-t_{f}\right) }$
for $t>t_{f}$, one obtains 
\begin{eqnarray}
\psi _{1}\left( x,t\right) =e^{-\beta
/2}\sum_{s_{1},s_{2}}i^{s_{1}}J_{s_{1}}\left( \theta /2\right)
I_{s_{2}}(\beta /2)  \nonumber \\
\exp \left\{ -i\omega _{q}\left[ s_{1}^{2}t_{f}+\left( s_{1}+s_{2}\right)
^{2}\left( t-t_{f}\right) \right] \right.   \nonumber \\
\left. +i\left( s_{1}+s_{2}\right) qx\right\} ,  \label{15}
\end{eqnarray}
where $I_{s}\left( x\right) $ is a modified Bessel function of order $s$.
From this expression, using an addition theorem for Bessel functions, one
finds that the atom density, $\rho \left( x,t\right) =\left| \psi \left(
x,t\right) \right| ^{2}$, is given by 
\begin{eqnarray}
\rho \left( x,t\right) =e^{-\beta }\sum_{s_{1},s_{2}}J_{s_{1}}\left\{ \theta
\sin \left[ \omega _{q}\left( s_{1}t+s_{2}\left( t-t_{f}\right) \right) %
\right] \right\}   \nonumber \\
\times I_{s_{2}}\left\{ \beta \cos \left[ \omega _{q}\left(
s_{1}+s_{2}\right) \left( t-t_{f}\right) \right] \right\} e^{i\left(
s_{1}+s_{2}\right) qx}.  \label{16}
\end{eqnarray}

Expressions (\ref{10}), (\ref{15}), and (\ref{16}) are convenient for
numerical evaluation of the atomic wave function and density. On the other
hand, they do not reveal in any transparent fashion the density peaks at the
focal plane $t=t_{f}$ and the high-order harmonic density patterns at the
fractional Talbot times 
\begin{equation}
t_{m}=t_{f}+T/m.  \label{17}
\end{equation}
Both of these features are more readily apparent if we follow alternative
approaches. First, instead of Eq. (\ref{10}), one can use approximate
expressions for the focal plane position and atom density profile in the
asymptotic limit of $\sqrt{\theta }\gg 1$ \cite{focusing}, 
\begin{mathletters}
\label{18}
\begin{eqnarray}
t_{f} &\sim &\omega _{q}^{-1}\theta ^{-1}\left( 1+1.27\theta ^{-1/2}\right) ,
\label{18a} \\
\rho \left( x,t_{f}\right)  &\sim &\sum_{s=-\infty }^{\infty }\frac{\sqrt{%
3\theta }}{\pi }  \nonumber \\
&&\times \left| f\left( 3^{1/4}\theta ^{3/4}\left( qx-2\pi s\right)
,-2.20\right) \right| ^{2},  \label{18b} \\
f\left( \tilde{x},\tilde{\omega}\right)  &=&\int_{-\infty }^{\infty }d\xi
\exp \left( -i\tilde{x}\xi +i\tilde{\omega}\xi ^{2}+i\xi ^{4}\right) .
\label{18c}
\end{eqnarray}
Equation (\ref{18b}) is valid within the small regions near the standing
wave nodes, $\left| x-s\lambda /2\right| \lesssim \lambda \theta ^{-3/4}.$

Second, one can also obtain alternative expressions for the atomic density
at the fractional Talbot times. Equation (\ref{16}) is valid for arbitrary $%
t>t_{f}.$\ However, if the time is restricted to be a rational fraction of
the Talbot time after the focus \cite{14}, 
\end{mathletters}
\begin{equation}
t=t_{f}+(\ell /m)T,  \label{191}
\end{equation}
for integer $\ell $ and $m,$ a derivation, similar to the one given in Ref. 
\cite{9}, leads to

\begin{mathletters}
\label{20}
\begin{eqnarray}
\psi _{1}\left( x,t\right) &=&\sum_{s=0}^{m-1}a_{s}\psi _{1}(x-s\lambda
/2m,t=t_{f}^{+}),  \label{20a} \\
a_{s} &=&(2im)^{-1}\sum_{s^{\prime }=0}^{\ell }\exp \left[ i\pi \left(
s+ms^{\prime }\right) ^{2}/2m\right] ,  \label{20b}
\end{eqnarray}
valid in the interval $0\leq x<\lambda /2$. The wave function at these times
(\ref{191}) appears as a superposition of $m$\ scaled images of the filtered
wave function at the focal time. In forming the atomic density, 
\end{mathletters}
\begin{equation}
\rho \left( x,t\right) =\left| \psi _{1}\left( x,t\right) \right| ^{2}\text{,%
}  \label{21}
\end{equation}
it is possible for different terms in the sum over $s$\ to interfere. (In
the absence of the optical mask ($\beta =0$), these interference terms cause
significant density oscillations that ruin the fractional self-image of the
atoms focused by the far-detuned field.) However, if the optical mask at $%
t_{f}$\ produces very narrow structures with an effective duty cycle much
less than unity, then the interference terms are unimportant and the density
reduces to 
\begin{equation}
\rho \left( x,t\right) =\sum_{s=0}^{m-1}\left| a_{s}\right| ^{2}\left| \psi
_{1}(x-s\lambda /2m,t=t_{f}^{+})\right| ^{2}\text{.}  \label{22}
\end{equation}

When valid, the spatial distribution (\ref{22}) becomes an exact $n$th-order
image of the density at $t_{f}$\ when all non-zero coefficients (\ref{20b})
have the same absolute value. This can occur at the fractional Talbot times 
\begin{equation}
t_{m}=t_{f}+(T/m),  \label{19}
\end{equation}
with $\ell =1$ in Eq. (\ref{191}). At these times,\ the coefficients become 
\begin{equation}
\left| a_{s}\right| =\left( 2/m\right) ^{1/2}\left\{ 
\begin{array}{c}
\left| \cos \left( m\pi /4\right) \right| ,\text{ for even }s \\ 
\left| \sin \left( m\pi /4\right) \right| ,\text{ for odd }s
\end{array}
\right.
\end{equation}
Three cases can be distinguished \cite{9}: 
\begin{mathletters}
\label{23}
\begin{eqnarray}
(a)\text{ }m &=&2n^{\prime }+1,  \label{23a} \\
(b)\text{ }m &=&2\left( 2n^{\prime }+1\right) ,  \label{23b} \\
(c)\text{ }m &=&4n^{\prime },  \label{23c}
\end{eqnarray}
where $n^{\prime }$ is a positive integer. In case (a), $\left| a_{s}\right|
^{2}=1/m$ for all $s,$ leading to atomic densities having period $\lambda /%
\left[ 2\left( 2n^{\prime }+1\right) \right] $. In case (b), $\left|
a_{s}\right| ^{2}=2/m$ for odd $s$ and $\left| a_{s}\right| ^{2}=0$ for even 
$s,$ leading to atomic densities having period $\lambda /\left[ 2\left(
2n^{\prime }+1\right) \right] $ but that are shifted by half a period from
those of case (a). In case (c), $\left| a_{s}\right| ^{2}=2/m$ for even $s$
and $\left| a_{s}\right| ^{2}=0$ for odd $s,$ leading to densities having
period $\lambda /4n^{\prime }$.

For the FTL scheme to produce high-quality, high-contrast fractional
self-images, it is sufficient that three conditions are met. First, the
focused atoms at $t_{f}$ must have a large density ($\rho \left(
x,t_{f}\right) \gg 1$) at the positions $x=2\pi s/q$. For $\theta \gg 1,$
the peak density can be estimated to grow as $3.83\theta ^{1/2}$ according
to an analysis of Eq. (\ref{18b})\cite{focusing}. Second, the focal spot
size $2w\approx 0.118\lambda \theta ^{-3/4}$\cite{focusing} should be
less than $\lambda /2m$ for case (a) and less than $\lambda /m$ for cases
(b) and (c). Thus, the effective duty cycle of the focused atom density over
the $\lambda /2$ period is 
\end{mathletters}
\begin{equation}
f\simeq 4w/\lambda \approx 0.237\theta ^{-3/4}\text{,}
\end{equation}
giving a maximum possible grating order on its own of 
\begin{equation}
n_{\max }\simeq f^{-1}\approx 4.22\theta ^{3/4}\text{.}
\end{equation}
Third, for Eq. (\ref{22}) to be valid so that interference terms in the
focused atom wave function do not degrade the fractional images at later
times, the optical mask should only pick out the density peaks and clip the
tails of the unavoidable background wave function between the peaks. If we
wish to use most of the focused atoms, the optical mask slit width or
opening size $\delta x$ should not be smaller than the focused atom spot
size $2w.$ From Eq. (\ref{13}), it follows that the transmission function
near the nodes for $\left| \chi _{2}\right| /\gamma \ll 1$ and $\Delta
_{2}=0 $ is $\eta (x)\simeq \exp \left[ -\beta \left( kx\right) ^{2}\right] $%
, from which we can estimate that the effective opening size $\delta x$ is 
\[
\delta x\simeq \left( \frac{\ln 2}{\beta }\right) ^{1/2}\frac{\lambda }{\pi }%
\approx 0.265\frac{\lambda }{\beta ^{1/2}}\text{.} 
\]
Setting $\delta x\gtrsim 2w,$ one finds the upper limit for $\beta $ 
\begin{equation}
\beta <5.04\theta ^{3/2}\text{.}  \label{231}
\end{equation}
To establish a lower limit for $\beta $, one must examine Eqs. (\ref{10}, 
\ref{13}, \ref{20}, \ref{22}) numerically for given values of $\theta $ and $%
m.$ We have found that a value of $\beta $ an order of magnitude smaller
than the upper limit given by Eq. (\ref{231}) can still give rise to high
contrast gratings.

\section{Discussion}

\begin{figure}[tb!]
\centering
\begin{minipage}{8.0cm}
\epsfxsize= 8 cm \epsfysize= 7.6 cm \epsfbox{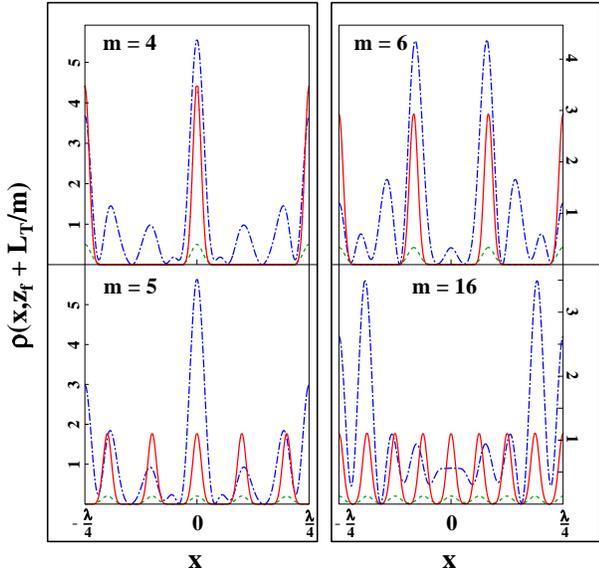}
\end{minipage}
\caption{Filtered Talbot lens production of high-order atomic density
patterns (solid lines) at distances $z_{f}+L_{T}/m$ for a mask parameter $%
\protect\beta =40$ in the weak field regime $\left| \protect\chi _{2}\right|
/\protect\gamma \ll 1\ll \protect\gamma \protect\tau .$ For comparison, we
show the spatial distributions of the metastable atomic density produced by
the optical mask alone at the fractional Talbot distances (dashed lines) and
by the focusing field alone at the fractional Talbot distances from the
focal plane (dot-dashed lines). For all plots, the focusing field area $%
\protect\theta $ is equal to $10.$}
\label{fig2}
\end{figure}

In Fig. \ref{fig2} we compare the atomic spatial distributions at the
fractional Talbot lengths when the focusing and masking fields each act
alone and when they are combined as detailed above. This plot was obtained
numerically using Eqs. (\ref{10}) and (\ref{20}) in (\ref{21}) for $\theta
=10$ and $\beta =40.$ [Alternatively, one can evaluate the density using Eq.
(\ref{16}). This equation must be used for distances following the focal
plane that are not a rational fraction of the Talbot time.] The effective
opening size of the optical mask (pulse 2) is larger than the spot size of
the focused atomic beam but still sufficiently narrow to filter out the
majority of background atoms in the tail of the wave function. From Eq. (\ref
{18a}) one finds $t_{f}\approx 0.140\omega _{q}^{-1},$ but we use the
more precise value $t_{f}\approx 0.146\omega _{q}^{-1},$ obtained
numerically in Ref. \cite{focusing}. The FTL produces high-contrast, higher
order atomic gratings at the fractional Talbot times. The mask alone also
produces such periodic structures but with severely reduced contrasts and
larger spot sizes. The focusing field acting alone produces narrow
structures at the fractional Talbot distances, but the density's overall
periodicity remains equal to $\lambda /2$ as opposed to forming a self-image
of reduced periodicity.

Within the confines of the Raman-Nath approximation, it might be difficult
to achieve a value $\beta =40$ if $\left| \chi _{2}\right| \ll \gamma $. For
example, when $\left| \chi _{2}\right| /\gamma \sim 0.1,$ one needs $\gamma
\tau $ to be as large as $4\times 10^{3}.$ For $\gamma /\omega _{q}\sim
10^{3}$ this implies values of $\omega _{q}\tau >1,$ which are inconsistent
with the Raman-Nath approximation. To avoid this difficulty, smaller values
of $\gamma \tau $ can be used with larger Rabi frequencies. Let us assume
that field 2, the standing wave mask, is a resonant, rectangular pulse with $%
\left| \chi _{2}\right| /\gamma \gg 1$ and that the adiabatic condition (\ref
{adiabatic}) is satisfied \cite{15}. Then, except in the vicinity of the
nodes, the transmission function (\ref{121}) is 
\begin{equation}
\eta \left( x\right) =\exp (i\left| \chi _{2}\tau \sin \left( kx\right)
\right| -\gamma \tau /2).  \label{25}
\end{equation}
From this expression, a value of $\gamma \tau \sim 3-4$ is enough to
attenuate the tails of the focused atoms with an accuracy of several
percent. Near the nodes, $\left| \chi _{2}\sin \left( kx\right) \right| \ll
\gamma $, and Eq. (\ref{14}) is still valid. As a result, large values of an
effective $\beta $ are possible for $\left| \chi _{2}\right| /\gamma \gg 1$
even if $\gamma \tau $ is on the order of unity.

\begin{figure}[tb!]
\centering
\begin{minipage}{8.0cm}
\epsfxsize= 8 cm \epsfysize= 7.6 cm \epsfbox{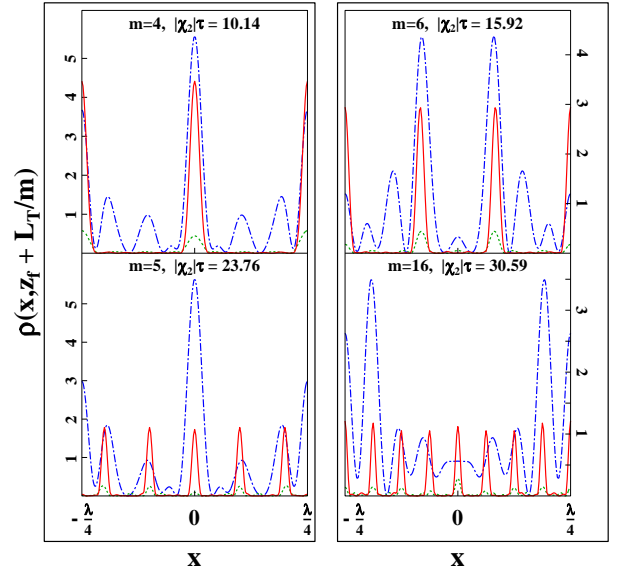}
\end{minipage}
\caption{The same type of density plots as those shown in Fig. \ref{fig2},
but for a moderate value of the upper state decay rate $\left( \protect\gamma
\protect\tau =4\right) $ and a large value of the optical mask Rabi
frequency $\left| \protect\chi _{2}/\protect\gamma \right| >1$ \ The
parameter $\left| \protect\chi _{2}\protect\tau \right| $ is chosen for each
plot by requiring that the ratio of the maximum and minimum amplitudes of
the peaks in the fractional density pattern are as close as possible to
unity. In each case this choice produces the best $n$th-order fractional
self-image with the filtered Talbot lens. For all plots, the focusing field
area $\protect\theta $ is equal to $10.$}
\label{fig3}
\end{figure}

In Fig. \ref{fig3} we plot the atom density profiles at fractional Talbot
distances with $\gamma \tau =4$ and $\theta =10.$ The plot has been obtained
by combining Eqs. (\ref{10}), (\ref{12}), (\ref{121}), and (\ref{20}) \cite
{16}. We have found that the amplitudes of the different peaks between $x=0$
and $x=\lambda /2$ in the fractional Talbot effect oscillate slightly as a
function of $\left| \chi _{2}\right| .$\ The origin of these Rabi-like
oscillations is an interference of the $m$\ terms in Eq. (\ref{20a}), each
of which has a different phase that depends on the Rabi frequency $\chi
_{2}. $\ For each $m$\ we have chosen a value of the parameter $\left| \chi
_{2}\right| \tau $\ that minimizes these oscillations to provide as pure a
fractional self-image as possible.

Chromatic aberration and atomic beam angular divergence degrade the focusing
and the Talbot effect. Chromatic aberration arises as a result of a
distribution of longitudinal velocities in the atomic beam. Recently, this
effect was considered in detail for thin lens atom focusing \cite{focusing}.
It was shown that, owing to the finite depth of focus, the focusing is not
seriously degraded for $\theta \approx 10$ until the spread in velocities
divided by the average velocity, $\Delta v/u,$ exceeds $0.1$. For larger
values of $\Delta v/u$, our numerical results indicate that high-order
atomic gratings are still produced, but a background density arises whose
amplitude starts to be comparable with the peaks of the fractional Talbot
spots.

Angular divergence in the beam provides a more severe restriction. If atoms
move at an angle $\phi $ relative to the $z$ axis, the atomic distribution
will be displaced by $\delta x\sim \phi L_{T}/m$ when the beam travels a
distance equal to $L_{T}/m.$ This displacement has to be smaller than the
width of the peak in the focused atomic distribution if we want to maintain
the beam flux and produce high-contrast, $m$th-order patterns. From Eq. (\ref
{18b}) one can deduce \cite{focusing} that the half-width of the
distribution is approximately equal to $0.06\lambda \theta ^{-3/4}$. From
this expression one finds that the angular divergence $\phi $ must be less
than $0.12\frac{\lambda _{dB}}{\lambda }m\theta ^{-3/4}$ for the FTL to
operate effectively. This requirement is similar to the one necessary for
the Talbot effect produced by microfabricated structures. We expect that the
application of modern sub-recoil laser cooling techniques or the use of Bose
condensates \cite{nist} will allow one to achieve these degrees of
longitudinal and transverse cooling.

In summary, we have presented a proposal for creating nanostructures using
both a focusing field and a masking field. The focusing field creates high
density focal spots that are transmitted by the masking field. At the focal
plane, the wave fronts entering the optical mask are coherent and fairly
flat. Consequently, the situation is similar to a collimated beam that
strikes the masking field. Owing to the Talbot effect, the spots are
refocused with reduced periods following the masking field at distances that
are integral fractions of the Talbot length. Using this technique, it should
be possible to produce high contrast structures having periods as small as $%
\lambda /20$.

\acknowledgments

J.L.C. is indebted to Prof. Tycho Sleator and the NYU Physics Department for
providing him with the Visiting Scholar appointment during which this work
was completed. This work is supported by the National Science Foundation
under Grants No. PHY-9414020 and PHY-9800981 and by the U.S. Army Research
Office under Grant No. DAAG55-97-0113 and AASERT No. DAAH04-96-0160.

\end{multicols}

\begin{references}
\bibitem{15}  A rectangular pulse is understood to have leading and falling
edges with durations $\delta \tau $ large enough so that the adiabatic
condition $\delta \tau \gg \min \left( \gamma ^{-1},\left| 2\chi _{2}\right|
^{-1},\left| \Delta _{2}\right| ^{-1}\right) $ is satisfied.
\bibitem{16}  Even for a large Rabi frequency $\left| \chi _{2}\right| ,$
one can still apply the Fourier method to derive an analog of Eq. (\ref{16})
for the atom density at an arbitrary time $t-t_{f}$. To do this, instead of
using an addition theorem for Bessel functions, one can use identity (121)
of Ref. \cite{9}, valid for any periodic transmission function. This
expression is not presented here.
\end{references}
\end{document}